\documentclass[prl, aps, twocolumn, preprintnumbers, showpacs, nofootinbib, superscriptaddress, notitlepage]{revtex4-1}
\usepackage{mathrsfs}
\usepackage{amsfonts}
\usepackage{amsmath}
\usepackage{slashed}
\usepackage{array}
\usepackage{verbatim}
\usepackage{epsfig}
\usepackage{graphicx}
\usepackage{color}
\usepackage{hyperref}
\usepackage[dvipsnames]{xcolor}
\usepackage{graphicx}   
\usepackage{color}    
\usepackage{slashed}   
\usepackage{verbatim}
\usepackage[normalem]{ulem}
\usepackage{rotating}   
\usepackage{multirow}  
\usepackage{inputenc}
\usepackage{bm}
\usepackage{booktabs} 
\usepackage{slashed}
\usepackage{multirow}
\usepackage{subcaption}
\usepackage{dsfont}

\allowdisplaybreaks
\usepackage{graphicx}
\usepackage[section]{placeins}
\usepackage{cleveref}
\usepackage{tikz}
\usepackage{natbib} 

\usepackage{caption}
\newcommand{\nn}{\nonumber}

\begin{document}

\title{Continuum-Limit HQET LCDAs from Lattice QCD for Tightening B Decay Uncertainties}

\author{\includegraphics[scale=0.10]{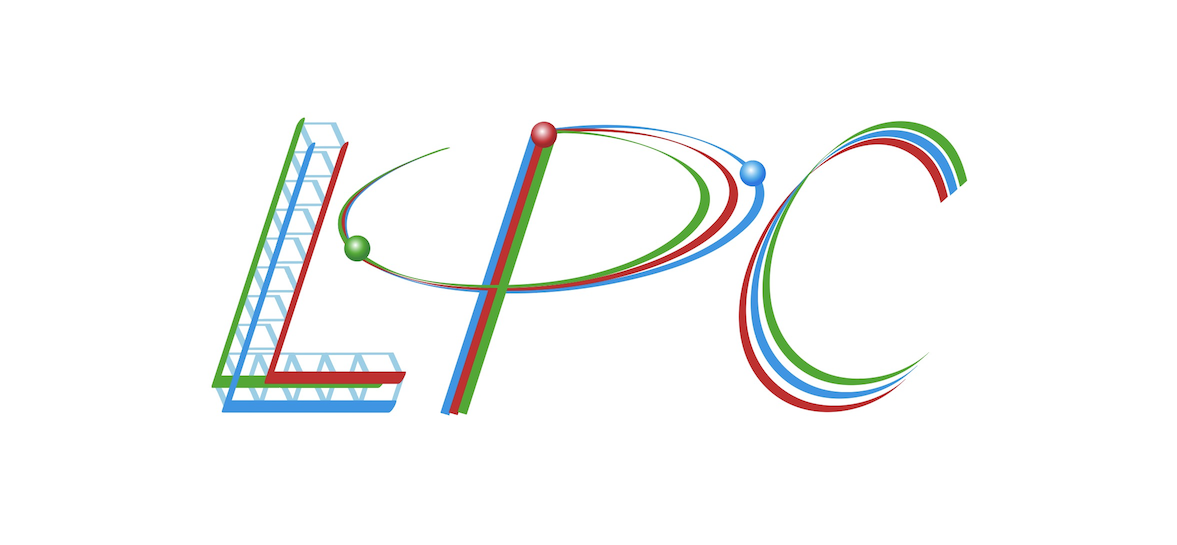}\\
Xue-Ying Han}
\affiliation{Institute of High Energy Physics, Chinese Academy of Sciences, Beijing 100049, China}
\affiliation{School of Physical Sciences, University of Chinese Academy of Sciences, Beijing 100049, China}

\author{Hao-Fei  Gao}
\affiliation{State Key Laboratory of Dark Matter Physics, Shanghai Key Laboratory for Particle Physics and Cosmology, Key Laboratory for Particle Astrophysics and Cosmology (MOE), School of Physics and Astronomy, Shanghai Jiao Tong University, Shanghai 200240, China}
\affiliation{Tsung-Dao Lee Institute, Shanghai Jiao Tong University, Shanghai 201210, China}

\author{Jun Hua}
\affiliation{Key Laboratory of Atomic and Subatomic Structure and Quantum Control (MOE), Guangdong
Basic Research Center of Excellence for Structure and Fundamental Interactions of Matter,
Institute of Quantum Matter, South China Normal University, Guangzhou 510006, China }
\affiliation{Guangdong-Hong Kong Joint Laboratory of Quantum Matter, Guangdong Provincial Key Laboratory of Nuclear Science, Southern Nuclear Science Computing Center, South China Normal University, Guangzhou 510006, China}

\author{Xiangdong Ji}
\affiliation{Maryland Center for Fundamental Physics, Department of Physics, University of Maryland, 4296 Stadium Dr., College Park, MD 20742, USA}
\affiliation{State Key Laboratory of Dark Matter Physics, Shanghai Key Laboratory for Particle Physics and Cosmology, Key Laboratory for Particle Astrophysics and Cosmology (MOE), School of Physics and Astronomy, Shanghai Jiao Tong University, Shanghai 200240, China}
\affiliation{Tsung-Dao Lee Institute, Shanghai Jiao Tong University, Shanghai 201210, China}

\author{Xiangyu Jiang}
\affiliation{CAS Key Laboratory of Theoretical Physics, Institute of Theoretical Physics, Chinese Academy of Sciences, Beijing 100190, China}

\author{Cai-Dian L\"u}
\affiliation{Institute of High Energy Physics, Chinese Academy of Sciences, Beijing 100049, China}
\affiliation{School of Physical Sciences, University of Chinese Academy of Sciences, Beijing 100049, China}

\author{Andreas Sch\"afer}
\affiliation{Institut f\"ur Theoretische Physik, Universit\"at Regensburg, D-93040 Regensburg, Germany}
\affiliation{Department of Physics, National Taiwan University, Taipei, Taiwan 106, China}

\author{Jin-Xin Tan}
\affiliation{State Key Laboratory of Dark Matter Physics, Shanghai Key Laboratory for Particle Physics and Cosmology, Key Laboratory for Particle Astrophysics and Cosmology (MOE), School of Physics and Astronomy, Shanghai Jiao Tong University, Shanghai 200240, China}
\affiliation{Tsung-Dao Lee Institute, Shanghai Jiao Tong University, Shanghai 201210, China}

\author{Ji-Hao Wang}
\affiliation{CAS Key Laboratory of Theoretical Physics, Institute of Theoretical Physics, Chinese Academy of Sciences, Beijing 100190, China}

\author{Wei Wang}
\email{Corresponding author: wei.wang@sjtu.edu.cn}
\affiliation{State Key Laboratory of Dark Matter Physics, Shanghai Key Laboratory for Particle Physics and Cosmology, Key Laboratory for Particle Astrophysics and Cosmology (MOE), School of Physics and Astronomy, Shanghai Jiao Tong University, Shanghai 200240, China}

\affiliation{Southern Center for Nuclear-Science Theory (SCNT), Institute of Modern Physics, Chinese Academy of Sciences, Huizhou 516000, Guangdong Province, China}

\author{Ji Xu}
\email{Corresponding author: xuji@lzu.edu.cn}
\affiliation{School of Nuclear Science and Technology, Lanzhou University, Lanzhou 730000, China}

\author{Yi-Bo Yang}
\affiliation{CAS Key Laboratory of Theoretical Physics, Institute of Theoretical Physics, Chinese Academy of Sciences, Beijing 100190, China}
\affiliation{School of Fundamental Physics and Mathematical Sciences, Hangzhou Institute for Advanced Study, UCAS, Hangzhou 310024, China}
\affiliation{International Centre for Theoretical Physics Asia-Pacific, Beijing/Hangzhou, China}
\affiliation{School of Physical Sciences, University of Chinese Academy of Sciences, 
Beijing 100049, China}

\author{Fu-Wei Zhang}
\affiliation{School of Physical Science and Technology, Inner Mongolia University, Hohhot 010021, China}

\author{Jian-Hui Zhang}
\affiliation{School of Science and Engineering, The Chinese University of Hong Kong, Shenzhen 518172, China}

\author{Jia-Lu Zhang}
\affiliation{Tsung-Dao Lee Institute, Shanghai Jiao Tong University, Shanghai 201210, China}
\affiliation{State Key Laboratory of Dark Matter Physics, Shanghai Key Laboratory for Particle Physics and Cosmology, Key Laboratory for Particle Astrophysics and Cosmology (MOE), School of Physics and Astronomy, Shanghai Jiao Tong University, Shanghai 200240, China}

\author{Mu-Hua Zhang}
\affiliation{State Key Laboratory of Dark Matter Physics, Shanghai Key Laboratory for Particle Physics and Cosmology, Key Laboratory for Particle Astrophysics and Cosmology (MOE), School of Physics and Astronomy, Shanghai Jiao Tong University, Shanghai 200240, China}
\affiliation{Tsung-Dao Lee Institute, Shanghai Jiao Tong University, Shanghai 201210, China}

\author{Qi-An Zhang }
\email{Corresponding author: zhangqa@buaa.edu.cn}
\affiliation{School of Physics, Beihang University, Beijing 102206, China}

\author{Shuai Zhao}
 \affiliation{School of Science, Tianjin University, Tianjin 300072, China}

\collaboration{Lattice Parton Collaboration}

\begin{abstract}
Heavy meson HQET light-cone distribution amplitudes (LCDAs) are critical for precision predictions of $B$ meson weak decays, but currently  are  one  of dominant theoretical uncertainties that obscure interpretations of $B$ anomalies and CP-violating measurements. Building on the established HQLaMET framework, supplemented by lattice QCD calculations of the OPE moments, we present a precise lattice QCD calculation of HQET LCDAs by employing multi-ensemble simulations for continuum and physical pion mass extrapolation, quantifying comprehensive systematic errors, and validating results through OPE moment cross-validation. Details of the lattice calculations are provided in a companion paper \cite{HeavymesonDA_long_paper}.
Our final results for key inverse moments (at $\mu=1$ GeV) are $\lambda_B=0.340(20)$ GeV and $\sigma_B^{(1)}=1.685(63)$, with the total uncertainty reduced by a factor of three relative to the previous analysis. These results can greatly reduce the uncertainty in the $B \to K^*$ form factors in the large-recoil region. This work resolves the long-standing bottleneck in first-principles predictions of heavy meson LCDAs, advancing precision flavor physics to new frontiers. 
\end{abstract}

\maketitle

{\it Introduction.} 
Weak decays of heavy mesons, particularly $B$ mesons, stand at the forefront of particle physics, serving as unparalleled probes for testing the standard model (SM) and searching for new physics beyond it \cite{Beneke:1999br,Lu:2000em,Ali:1999mm}. Processes like $B \to K^*\ell^+\ell^-$ (flavor-changing neutral currents)  are exquisitely sensitive to deviations from SM predictions: even small discrepancies (e.g., the ``$B$ anomaly" in $R_{K^{(*)}} = \Gamma(B \to K^{(*)}\mu^+\mu^-)/\Gamma(B \to K^{(*)}e^+e^-)$ \cite{Ali:1999mm,LHCb:2021trn}) at low $q^2$ could signal new particles or interactions. However, unlocking this potential requires theoretical predictions with precision comparable to experimental measurements, an achievement long blocked by a critical bottleneck: heavy meson light-cone distribution amplitudes (LCDAs) in heavy quark effective theory (HQET) \cite{Grozin:1996pq,Braun:2003wx}. Theoretical  prediction   depends on heavy meson LCDAs from two aspects: first in QCD factorization, the decay amplitudes contain hard-scattering contributions, in which the HQET LCDAs are indispensable; secondly, in the calculation of form factors at low $q^2$, the light-cone sum rules (LCSRs) \cite{DeFazio:2005dx,Khodjamirian:2006st,Wang:2015vgv,Lu:2018cfc,Gao:2019lta,Cui:2022zwm,Gao:2024vql,Huang:2025jsa,Li:2025mhq} are often used and require the LCDAs. 

HQET LCDAs encode the nonperturbative momentum partitioning between the heavy quark and light (anti)quark in the meson. Unlike light-meson LCDAs, which can be constrained by lattice QCD and experiment \cite{LatticeParton:2022zqc,BaBar:2009rrj,Belle:2012wwz}, heavy meson HQET LCDAs have resisted first-principles determination for more than three decades. 
The difficulty is fundamental: light-cone correlations are not directly accessible in Euclidean lattice QCD, while cusp divergences invalidate the conventional local-operator expansion and obstruct the standard moment-based strategy \cite{Ji:2013dva,Braun:2003wx,Korchemskaya:1992je}. As a result, phenomenological analyses have long relied on ad hoc model parametrizations with poorly constrained shape parameters \cite{Lee:2005gza,Braun:2003wx,Grozin:1996pq}. This model dependence induces substantial hadronic uncertainties in heavy meson decay amplitudes and form factors; in important low-$q^2$ channels such as $B\to K^*$, the resulting uncertainties can exceed the precision already reached by modern flavor experiments \cite{Gao:2019lta}. Removing this long-standing LCDA bottleneck is thus not merely a technical refinement, but a prerequisite for precision heavy flavor phenomenology and meaningful new physics searches.

A decisive advance toward resolving this long-standing problem was recently achieved within the HQLaMET framework \cite{Han:2024fkr,LatticeParton:2024zko}, a sequential matching program that combines Large Momentum Effective Theory (LaMET) \cite{Ji:2013dva,Ji:2014gla,Ji:2020ect,Cichy:2018mum} with boosted Heavy-Quark Effective Theory (HQET) \cite{Beneke:2023nmj}. By exploiting the hierarchical scale ordering $P^z \gg m_H \gg \Lambda_{\rm QCD}$, HQLaMET enables lattice QCD access to boosted equal-time QCD correlators, which can then be systematically matched onto the HQET LCDAs. This established, for the first time, a model-independent and first-principles route to heavy meson LCDAs \cite{LatticeParton:2024zko,Han:2024fkr}. However, the existing studies remained at the proof-of-concept stage, being limited to a single lattice spacing and lacking full control over statistical and systematic uncertainties. This has so far prevented the method from delivering phenomenological precision commensurate with modern flavor experiments: while Belle II and LHCb already measure many exclusive $B$ decays at the sub-10\% level, LCDA-induced theory uncertainties still exceed 20\%, blurring the interpretation of precision flavor observables in terms of hadronic QCD dynamics versus possible new-physics effects.
  
In this Letter, together with the companion paper \cite{HeavymesonDA_long_paper}, we elevate this program from proof-of-concept to precision determination. Using multi-ensemble lattice simulations, continuum and physical-pion-mass extrapolations, a comprehensive treatment of statistical and systematic uncertainties, and nontrivial cross-checks from OPE moments, we present the most precise determination of heavy meson HQET LCDAs to date. Relative to the previous calculation \cite{Han:2024fkr}, the total uncertainty is reduced by about a factor of three. More importantly, this work turns heavy meson HQET LCDAs from model assumptions into systematically improvable nonperturbative inputs, significantly sharpening the theoretical basis of exclusive $B$-decay phenomenology. Beyond its immediate phenomenological impact, it establishes a practical framework for accessing heavy-hadron light-cone structure from lattice QCD.

{\it Method.}
Our analysis is based on the HQLaMET sequential matching framework \cite{Han:2024fkr,LatticeParton:2024zko,HeavymesonDA_long_paper}, which exploits the scale hierarchy
\begin{align}
\Lambda_{\rm QCD}\ll m_H\sim m_Q \ll P^z ,
\end{align}
so that the inaccessible heavy meson HQET LCDA can be related, through two controlled perturbative matching steps, to equal-time Euclidean correlators directly computable in lattice QCD.

The target quantity is the leading-twist HQET LCDA,
\begin{align}
	i\tilde{f}_H &m_H \varphi ^+(\omega,\mu) = \int\frac{dt}{2\pi}e^{it\omega n_+\cdot v}\nonumber\\
&\times\langle0\left|\bar{q}_s(tn_+)n\!\!\!\slash_+\gamma_5 W_c(tn_+,0)h_v(0) \right|H(v)\rangle, 
\label{eq:definition_of_HQETLCDA}
\end{align}
which governs the light-cone momentum distribution of the light degrees of freedom inside the heavy meson. Since this light-cone correlator is not directly accessible in Euclidean space, we instead start from the renormalized quasi-distribution amplitude (quasi-DA) from a spatial equal-time correlator of a highly boosted heavy meson,
\begin{align}
	\tilde{\phi}(x,P^z) =& \int\frac{dz}{2\pi}e^{-ixP^zz} \nn\\
	&\times \frac{\langle0|\bar{q}(z\hat{n}_z)\Gamma W_c(z\hat{n}_z,0)Q(0)|H(P^z)\rangle}{\langle0|\bar{q}(0)\Gamma Q(0)|H(P^z)\rangle}_R,
\label{eq:def_of_quasiDA}
\end{align}
with $\hat{n}_z=(0,0,0,1)$. For pseudoscalar mesons we choose $\Gamma=\gamma^z\gamma_5$, which matches onto the leading-twist light-cone amplitude while avoiding additional operator-mixing effects on the lattice \cite{Liu:2018tox,Liu:2019urm}.

In the large $P^z$ limit, the quasi-DA shares the same infrared physics with the QCD LCDA, so they can connected through the standard LaMET matching, \cite{Liu:2018tox,Xu:2018mpf}
\begin{align}
	\phi(y,\mu)=&\int_{-\infty}^{+\infty}dx~ \mathcal{C}\left(x,y,\frac{\mu}{P^z}\right)\tilde{\phi}(x,P^z) \nn\\
	&\quad +\mathcal{O}\left(\frac{m_H^2}{(P^z)^2}, \frac{\Lambda_{\mathrm{QCD}}^2}{(yP^z, \bar{y}P^z)^2} \right),
\label{eq:LaMET_match}
\end{align}
where $y$ is the light-quark momentum fraction in the QCD LCDA. Physically, this step integrates out the largest scale $P^z$ and converts the lattice-accessible equal-time correlator into the corresponding light-cone distribution in full QCD.

The second step is specific to heavy mesons and integrates out the heavy-quark scale. In the heavy-quark limit, the QCD LCDA is highly asymmetric and concentrated near the endpoint region $y\sim \Lambda_{\rm QCD}/m_H$. Introducing the natural HQET variable $\omega\equiv y m_H$, one separates the light-cone distribution into a nonperturbative peak region with $\omega\sim\mathcal{O}(\Lambda_{\rm QCD})$ and a hard tail at $\omega\sim\mathcal{O}(m_H)$. The peak region of HQET LCDA is governed by its QCD counterpart through an effective theory, while the difference between them is encoded in a perturbative jet function \cite{Beneke:2023nmj} (see also \cite{Ishaq:2019dst,Zhao:2019elu}),
\begin{align}
	\varphi^+_{\rm peak}(\omega,\mu) = \frac{1}{m_H}\frac{f_H}{\tilde{f}_H} \frac{1}{\mathcal{J}_{\rm peak}} \phi(y,\mu),
  \label{eq:HQET_matching_peak}
\end{align}
while the large-$\omega$ tail is perturbatively calculable \cite{Lee:2005gza}. 
This two-step factorization cleanly separates the boost scale $P^z$ from the heavy-quark scale $m_H$, thereby turning the heavy meson LCDAs into a systematically improvable quantity in lattice QCD.

{\it Results.}
The lattice input in this Letter is obtained from $N_f=2+1$ gauge ensembles with lattice spacings ranging from $a\simeq 0.105$ to $0.052~\mathrm{fm}$ and pion masses from physical to about $320~\mathrm{MeV}$ \cite{CLQCD:2023sdb,CLQCD:2024yyn}. We take the $D$ meson as the representative heavy meson, for which the hierarchy $\Lambda_{\rm QCD}\ll m_H\ll P^z$ can be approached with currently accessible lattice momenta. The LaMET reconstruction of the QCD LCDA is performed on the finer ensembles, with boost momenta up to $P^z\simeq 3.5~\mathrm{GeV}$ on the finest lattice, enabling controlled continuum, chiral, and $P^z\to\infty$ extrapolations. To improve the signal quality of boosted nonlocal correlators, we employ momentum-smeared sources and HYP-smeared Wilson lines \cite{Tan:2025ofx}. Further details of the lattice setup and analysis are given in the companion paper \cite{HeavymesonDA_long_paper}.

In parallel, we compute the lowest moments of the QCD LCDA from local lattice operators on all ensembles, providing an independent OPE benchmark for the LaMET reconstruction. The lattice-determined QCD LCDA in the soft region, $\omega\equiv y m_H\sim\mathcal{O}(\Lambda_{\rm QCD})$, is then used to constrain the nonperturbative peak of $\varphi^+(\omega,\mu)$, while the large-$\omega$ tail is obtained perturbatively from the HQET matching. A model-independent parametrization is finally used to connect the peak and tail smoothly, yielding the HQET LCDA together with a controlled assessment of statistical and systematic uncertainties.

Compared with the previous proof-of-concept studies, the present work upgrades the determination of both QCD and HQET LCDAs of heavy meson to the precision level in three essential aspects. First, we extend the calculation to multi dynamical fermion ensembles, which enables controlled continuum, physical-pion-mass, and finite-$P^z$ extrapolations. Second, we carry out a comprehensive analysis of systematic uncertainties, with all major sources explicitly quantified using standard lattice-QCD procedures. Third, we perform an independent determination of the lowest LCDA moments through the lattice OPE and use it as a nontrivial cross-check of the LaMET reconstruction. Together, these advances substantially improve the reliability of the final LCDA determination and provide the level of uncertainty control required for precision heavy-flavor phenomenology.

\begin{figure}[tbp]
  \centering
  \includegraphics[width=1.0\linewidth]{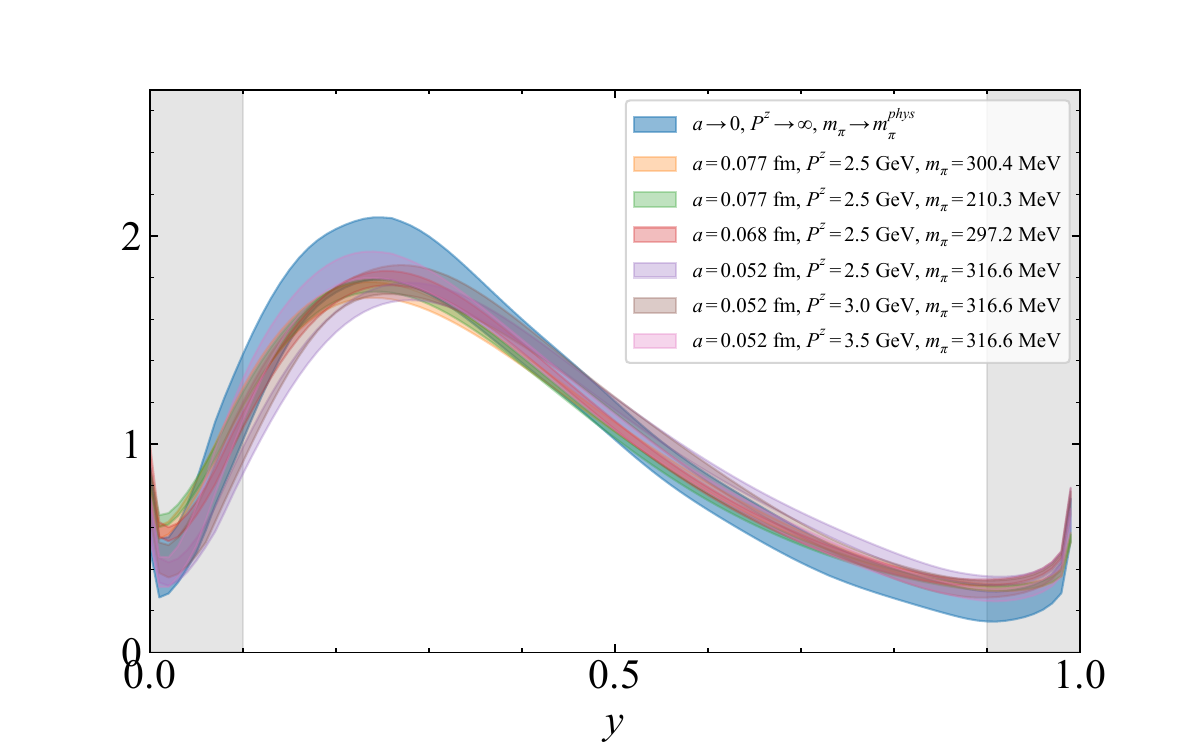}
  \caption{$D$-meson QCD LCDA $\phi(y,\mu=m_D)$ obtained from LaMET matching on different ensembles and at different momenta. The blue band shows the final result after extrapolating to $a\to0$, $P^z\to\infty$, and $m_\pi\to m_\pi^{\rm phys}$. Only statistical uncertainties are included in the bands.}
  \label{fig:qcd_lcda_ensembles}
\end{figure}

\begin{figure}[tbp]
\centering
\includegraphics[width=0.9\linewidth]{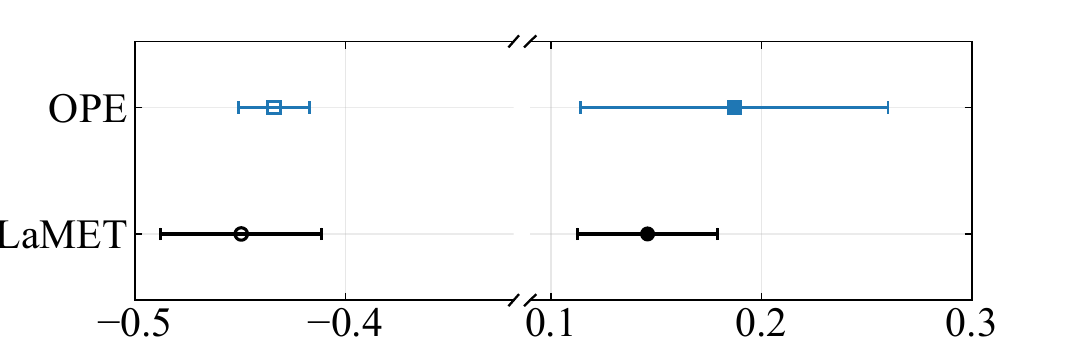}
\caption{Comparison of the Gegenbauer moments $a_1$ (left) and $a_2$ (right) of the $D$-meson QCD LCDA at $\mu=m_D$ from LaMET (black) and OPE (blue).}
\label{fig:moment_compare_LaMET_OPE}
\end{figure}

We first present the QCD LCDA in Fig.~\ref{fig:qcd_lcda_ensembles}. 
The results from different ensembles and boost momenta exhibit a consistent behavior and yield a stable continuum--chiral--infinite-momentum extrapolation. 
As an independent validation of the LaMET reconstruction, we compare its moments with those directly computed from the lattice OPE. 
The latter gives $\langle \xi \rangle_{\mu=m_D}=-0.260(10)$ and $\langle \xi^2 \rangle_{\mu=m_D}=0.262(23)$, with $\xi=2y-1$. 
Using the standard relations \cite{Efremov:1979qk,Lepage:1980fj}, these results correspond to $a_1^{\rm OPE}(\mu)=-0.434(17)$ and $a_2^{\rm OPE}(\mu)=0.183(73)$, in agreement with the LaMET determinations $a_1^{\rm LaMET}(\mu)=-0.449(38)$ and $a_2^{\rm LaMET}(\mu)=0.146(33)$, as shown in Fig.~\ref{fig:moment_compare_LaMET_OPE}. 
This agreement provides a nontrivial check that finite-$P^z$ effects, higher-power corrections, and perturbative matching uncertainties are under control within the present precision.

As phenomenological application of our precision lattice predictions, we predict the rare exclusive radiative decays $W\to D\gamma$ and $W\to B\gamma$, two benchmark observables directly sensitive to heavy meson nonperturbative dynamics. Using our first-principles QCD and HQET LCDAs as input, we obtain
\begin{align}
  \begin{aligned}
 {\cal B}(W\to D\gamma)=(9.6\pm0.9_{I}{}^{+0.5}_{-0.6}{}_{\mu}\pm2.2_{L})\times 10^{-10},
  \end{aligned}
  \label{eq:WtoDgamma}
\end{align}
and
\begin{align}
  \begin{aligned}
{\cal B}(W\to B\gamma)=(2.02\pm0.17_{I}{}^{+0.18}_{-0.21}{}_{\mu}\pm0.29_{L})\times 10^{-12},
  \end{aligned}
  \label{eq:WtoBgamma}
\end{align}
where the quoted uncertainties arise from input parameters ($I$), scale variation ($\mu$), and lattice data ($L$). These predictions establish precise SM benchmarks for future experimental studies and highlight the broader phenomenological impact of a systematically controlled lattice determination of heavy-meson LCDAs.

\begin{figure}[tbp]
  \centering
  \includegraphics[width=1.0\linewidth]{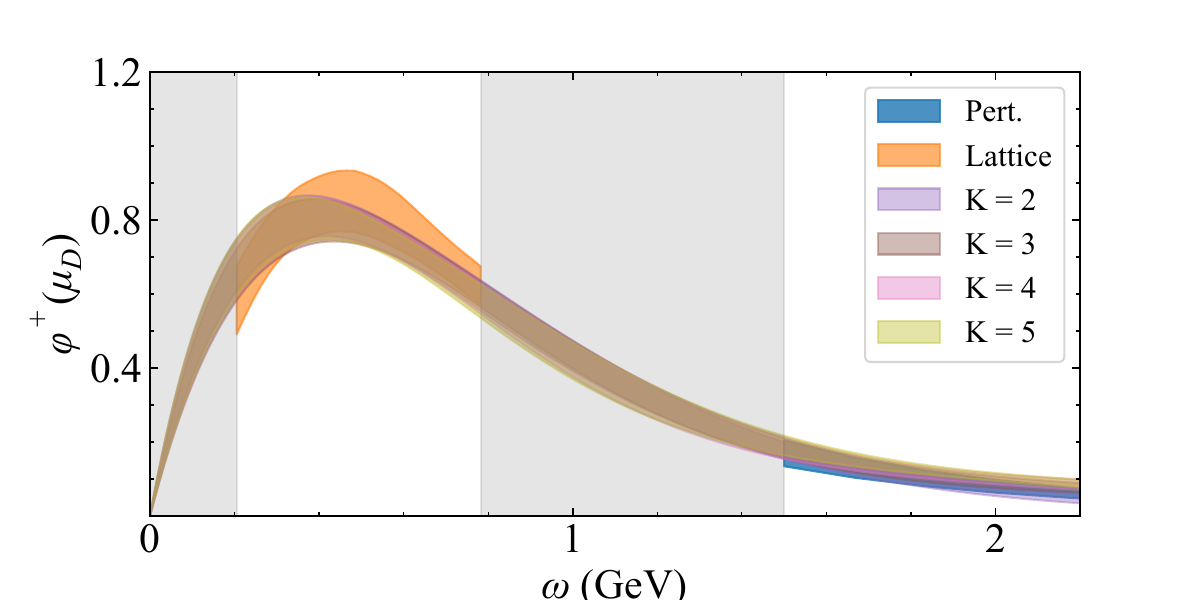}
  \caption{Reconstruction of the HQET LCDA $\varphi^+(\omega,\mu)$ using the Laguerre-polynomial parametrization in Eq.~\eqref{eq:HQETLCDAsmallomegaexpansion2}. The bands correspond to fits with truncation orders $K=2,3,4,5$, constrained by the lattice-determined nonperturbative peak region and the perturbative large-$\omega$ tail.}
  \label{fig:HQET_LCDA_param}
\end{figure}

Beyond the benchmark of  QCD LCDAs discussed above, our analysis provides first-principles constraints on the HQET LCDA itself. The resulting distribution $\varphi^+(\omega,\mu=m_D)$ is shown in Fig.~\ref{fig:HQET_LCDA_param}, where $\omega=y\,m_D$ denotes the light-quark energy in the heavy-quark rest frame. From first principles, however, the distribution is not determined point by point over the entire $\omega$ range. Instead, the nonperturbative peak region at $\omega=\mathcal{O}(\Lambda_{\rm QCD})$ is constrained by the lattice QCD input through the HQLaMET matching, while the large-$\omega$ tail is fixed by perturbation theory. These two regions, corresponding to the unshaded parts of Fig.~\ref{fig:HQET_LCDA_param}, are therefore under direct first-principles control.

To reconstruct the full distribution of $\varphi^+(\omega,\mu)$, we introduce the parametrization of Eq.~\eqref{eq:HQETLCDAsmallomegaexpansion2} \cite{Feldmann:2022uok},
\begin{align}
	\varphi^+(\omega,\mu)= \frac{\omega\, e^{-\omega / \omega_0}}{\omega_0^2}
	\sum_{k=0}^K \frac{b_k(\mu)}{1+k}\,L_k^{(1)}\!\left(\frac{2\omega}{\omega_0}\right),
	\label{eq:HQETLCDAsmallomegaexpansion2}
\end{align}
and determine its parameters by requiring consistency with the first-principles information in the peak and tail regions. In this way, the parametrization provides a smooth interpolation across the intermediate region, where no direct first-principles determination is available. As shown in Fig.~\ref{fig:HQET_LCDA_param}, the reconstructed distributions exhibit good convergence and stability with respect to the truncation order $K$.

\begin{figure}[http]
  \centering
  \includegraphics[width=1.00\linewidth]{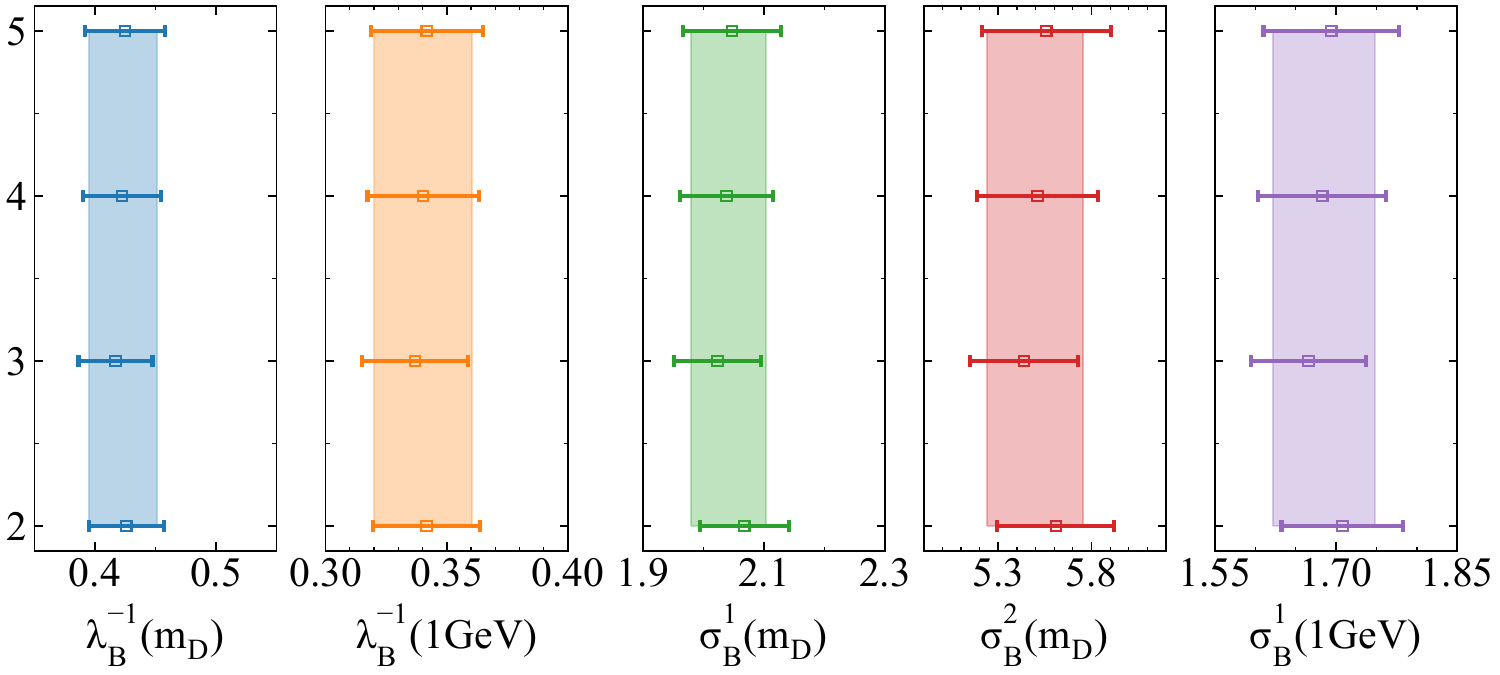}
  \caption{The first inverse moment $\lambda_B$ and inverse-logarithmic moments $\sigma_B^{(1,2)}$ of the HQET LCDA at $\mu=m_D$ and $\mu=1~\mathrm{GeV}$. The bands are obtained from constant fits to the results for truncation orders $K=2,3,4,5$.}
  \label{fig:HQLaMET_inverse_moment}
\end{figure}

With the full distribution of HQET LCDA, we can further predict the phenomenologically important inverse moments. In particular, we extract the first inverse moment and the first two inverse-logarithmic moments. As shown in Fig.~\ref{fig:HQLaMET_inverse_moment} and summarized in Table~\ref{tab:lambdaB_sigmaB_summary}, the extracted values are stable against the truncation order $K$ in Eq.~\eqref{eq:HQETLCDAsmallomegaexpansion2}. Notably, despite the inclusion of a much more complete treatment of systematic uncertainties than in our exploratory study \cite{LatticeParton:2024zko},the present calculation achieves substantially improved precision, while remaining consistent with the previous determination.

\begin{table*}[http]
  \centering
  \renewcommand{\arraystretch}{1.8}
  \setlength{\tabcolsep}{2.5mm}
  \begin{tabular}{l l c c c}
    \hline
    $\mu$ & Reference (Method) & $\lambda_B~(\mathrm{GeV})$ & $\sigma_B^{(1)}$ & $\sigma_B^{(2)}$ \\
    \hline
    $m_D$  & This work & 0.423(28) & 2.041(62) & 5.50(26) \\
           & Ref.~\cite{LatticeParton:2024zko} (LQCD) & 0.420(71) & 2.17(16) & 6.33(80) \\
    \hline
    $1~\mathrm{GeV}$    & This work & 0.340(20) & 1.685(63) & -- \\
           & Ref.~\cite{LatticeParton:2024zko} (LQCD) & 0.376(63) & 1.66(13) & -- \\
           & Ref.~\cite{Belle:2018jqd} (Experiment) & $>0.24$ & -- & -- \\
           & Ref.~\cite{Gao:2019lta} (QCD sum rule)  & $0.343^{+0.064}_{-0.079}$ & 1.4(4) & -- \\
           & Ref.~\cite{Braun:2003wx} (QCD sum rule)  & 0.46(11) & 1.4(4) & -- \\
           & Ref.~\cite{Khodjamirian:2020hob} (QCD sum rule) & 0.383(153) & -- & -- \\
           & Ref.~\cite{Lee:2005gza} (OPE)  & 0.48(11) & 1.6(2) & -- \\
           & Ref.~\cite{Grozin:1996pq} (Asymptotic behavior)  & 0.35(15) & -- & -- \\
           & Ref.~\cite{Mandal:2023lhp} (Global Fit) & 0.338(68) & -- & -- \\
    \hline
  \end{tabular}
  \caption{Summary of the first inverse moment $\lambda_B$ and inverse-logarithmic moments $\sigma_B^{(n)}$ from this Letter and previous determinations.}
  \label{tab:lambdaB_sigmaB_summary}
\end{table*}

\begin{figure}[http]
\centering
\includegraphics[width=0.9\linewidth]{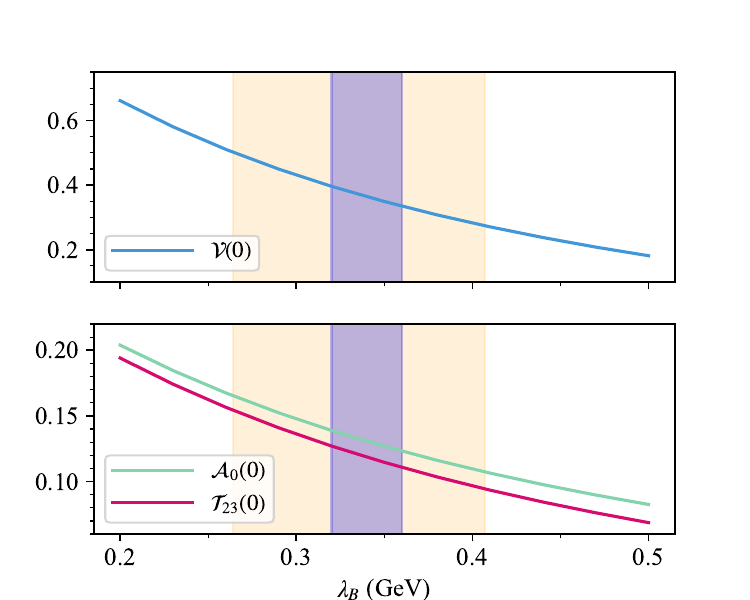}
\caption{Dependence of the $B\to K^*$ form factors on the inverse moment $\lambda_B$ of the heavy-meson LCDA. The yellow band uses the original input for $\lambda_B$ from Ref.~\cite{Gao:2019lta}, while purple band shows the reduced uncertainty obtained with our lattice determination.}
  \label{fig:B_Kstar_formfactors}
\end{figure}

Using the analytic LCSR calculation in Ref.~\cite{Gao:2019lta}, we use our result for $\lambda_B$ and recompute the $B\to K^*$ form factors. In Ref.~\cite{Gao:2019lta}, the input $\lambda_B=(0.343^{+0.064}_{-0.070})\,\mathrm{GeV}$ led to
\begin{align}
  V(0) =~& 0.359^{+0.141}_{-0.085}, \quad
  A_0(0) = 0.129^{+0.035}_{-0.021}, \nonumber\\
  &T_{23}(0) = 0.116^{+0.036}_{-0.022},
  \label{eq:BKstar_FF_GLSWW}
\end{align}
whereas replacing this input by our evolved result at $\mu=1~\mathrm{GeV}$, $\lambda_B=(0.340\pm0.020)\,\mathrm{GeV}$, yields
\begin{align}
  V(0) =~& 0.362^{+0.034}_{-0.029}, \quad
  A_0(0) = 0.131^{+0.008}_{-0.008}, \nonumber\\
  &T_{23}(0) = 0.119^{+0.008}_{-0.008}.
  \label{eq:BKstar_FF_ThisWork}
\end{align}
The comparison, displayed in Fig.~\ref{fig:B_Kstar_formfactors}, shows a substantial reduction of the $\lambda_B$-induced uncertainty across all three form factors. In particular, the upper uncertainty of $V(0)$ is reduced by about a factor of five, from ${}^{+0.141}_{-0.085}$ to ${}^{+0.034}_{-0.029}$, while the uncertainties of $A_0(0)$ and $T_{23}(0)$ shrink from ${}^{+0.035}_{-0.021}$ to ${}^{+0.008}_{-0.008}$ and from ${}^{+0.036}_{-0.022}$ to ${}^{+0.008}_{-0.008}$, respectively. This illustrates directly how a systematically controlled lattice determination of heavy-meson LCDAs can sharpen phenomenological predictions and strengthen precision tests in heavy-flavor physics.

{\it Summary and Prospect.} 
Heavy meson HQET LCDA is indispensable nonperturbative inputs for precision studies of weak $B$-meson decays, yet their long-standing model dependence has remained a major limitation in exclusive heavy-flavor phenomenology. Building on the HQLaMET framework, we have presented a precision lattice QCD determination of the heavy meson HQET LCDA and substantially advanced the subject beyond the proof-of-concept stage. The key improvements include multi-ensemble simulations enabling controlled continuum and physical-pion-mass extrapolations, a comprehensive treatment of statistical and systematic uncertainties, and an independent validation through OPE moments. 

Our final results for the most important inverse moments at $\mu=1~\mathrm{GeV}$ are $\lambda_B=0.340(20)~\mathrm{GeV}$ and $\sigma_B^{(1)}=1.685(63)$, with the overall uncertainty reduced by about a factor of three relative to our exploratory study. These results reduce the LCDA-induced uncertainties in $B \to K^*$ form factors by nearly a factor of two compared to Ref.~\cite{Gao:2019lta}, enabling definitive tests of the SM and robust interpretations of $B$ anomalies.

More broadly, this work transforms heavy meson HQET LCDA from model-dependent assumptions into systematically improvable quantities rooted in first-principles QCD. It therefore  provides a realistic method to remove a bottleneck in precision heavy-flavor physics and establishes a practical first-principles framework for bringing heavy meson light-cone structure under quantitative theoretical control. Future lattice simulations on finer lattices can yield more precise results, as they access larger momentum scales and heavier quark masses, thereby reducing power corrections~\cite{Han:2024cht,Guo:2025obm,Deng:2024dkd,Wang:2025uap}.

{\it Acknowledgement.} 
We thank Yu-Ming Wang for valuable discussions, and providing the error budget for $B\to\pi$ form factors and his code to calculate the $B\to V$ form factors in $B$ meson LCSRs. We thank the CLQCD collaborations for providing us the gauge configurations with dynamical fermions~\cite{CLQCD:2023sdb,CLQCD:2024yyn}, which are generated on the HPC Cluster of ITP-CAS, the Southern Nuclear Science Computing Center(SNSC), the Siyuan-1 cluster supported by the Center for High Performance Computing at Shanghai Jiao Tong University, and the Dongjiang Yuan Intelligent Computing Center.  This work is supported in part by National Natural Science Foundation of China under grants No.12125503, 12305103, 12375069, 12375080, 12525504, 12435002, 12293060, 12293062, 12275277, 12435004 and 12447101. CDL is also is partly supported by the National Key Research and Development Program of China (2023YFA1606000). YBY is also supported in part by National Key R\&D Program of China No.2024YFE0109800, and the Strategic Priority Research Program of Chinese Academy of Sciences, Grant No. YSBR-101. QAZ is also supported by the Fundamental Research Funds for the Central Universities. JHZ is also supported by Shenzhen Fundamental Research Grant No. JCYJ20250604141224032, the Ministry of Science and Technology of China under Grant No. 2024YFA1611004, and by CUHK-Shenzhen under grant No. UDF01002851.


\begin{thebibliography}{99}

\bibitem{HeavymesonDA_long_paper} 
Lattice Parton Collaboration (LPC), ``Determination of heavy meson light-cone distribution amplitudes: theoretical framework and lattice simulations," 
companion paper, submitted (2026).

\bibitem{Beneke:1999br}
M.~Beneke, G.~Buchalla, M.~Neubert and C.~T.~Sachrajda,
Phys. Rev. Lett. \textbf{83}, 1914-1917 (1999)
doi:10.1103/PhysRevLett.83.1914
[arXiv:hep-ph/9905312 [hep-ph]].

\bibitem{Lu:2000em}
C.~D.~Lu, K.~Ukai and M.~Z.~Yang,
Phys. Rev. D \textbf{63}, 074009 (2001)
doi:10.1103/PhysRevD.63.074009
[arXiv:hep-ph/0004213 [hep-ph]].

\bibitem{Ali:1999mm}
A.~Ali, P.~Ball, L.~T.~Handoko and G.~Hiller,
Phys. Rev. D \textbf{61}, 074024 (2000)
doi:10.1103/PhysRevD.61.074024
[arXiv:hep-ph/9910221 [hep-ph]].

\bibitem{LHCb:2021trn}
R.~Aaij \textit{et al.} [LHCb],
Nature Phys. \textbf{18}, no.3, 277-282 (2022)
doi:10.1038/s41567-023-02095-3
[arXiv:2103.11769 [hep-ex]].

\bibitem{Grozin:1996pq}
A.~G.~Grozin and M.~Neubert,
Phys. Rev. D \textbf{55}, 272-290 (1997)
doi:10.1103/PhysRevD.55.272
[arXiv:hep-ph/9607366 [hep-ph]].

\bibitem{Braun:2003wx}
V.~M.~Braun, D.~Y.~Ivanov and G.~P.~Korchemsky,
Phys. Rev. D \textbf{69}, 034014 (2004)
doi:10.1103/PhysRevD.69.034014
[arXiv:hep-ph/0309330 [hep-ph]].

\bibitem{DeFazio:2005dx}
F.~De Fazio, T.~Feldmann and T.~Hurth,
Nucl. Phys. B \textbf{733}, 1-30 (2006)
[erratum: Nucl. Phys. B \textbf{800}, 405 (2008)]
doi:10.1016/j.nuclphysb.2008.03.022
[arXiv:hep-ph/0504088 [hep-ph]].

\bibitem{Khodjamirian:2006st}
A.~Khodjamirian, T.~Mannel and N.~Offen,
Phys. Rev. D \textbf{75}, 054013 (2007)
doi:10.1103/PhysRevD.75.054013
[arXiv:hep-ph/0611193 [hep-ph]].

\bibitem{Wang:2015vgv}
Y.~M.~Wang and Y.~L.~Shen,
Nucl. Phys. B \textbf{898}, 563-604 (2015)
doi:10.1016/j.nuclphysb.2015.07.016
[arXiv:1506.00667 [hep-ph]].

\bibitem{Lu:2018cfc}
C.~D.~L{\"u}, Y.~L.~Shen, Y.~M.~Wang and Y.~B.~Wei,
JHEP \textbf{01}, 024 (2019)
doi:10.1007/JHEP01(2019)024
[arXiv:1810.00819 [hep-ph]].

\bibitem{Gao:2019lta}
J.~Gao, C.~D.~L{\"u}, Y.~L.~Shen, Y.~M.~Wang and Y.~B.~Wei,
Phys. Rev. D \textbf{101}, no.7, 074035 (2020)
doi:10.1103/PhysRevD.101.074035
[arXiv:1907.11092 [hep-ph]].

\bibitem{Cui:2022zwm}
B.~Y.~Cui, Y.~K.~Huang, Y.~L.~Shen, C.~Wang and Y.~M.~Wang,
JHEP \textbf{03}, 140 (2023)
doi:10.1007/JHEP03(2023)140
[arXiv:2212.11624 [hep-ph]].

\bibitem{Gao:2024vql}
J.~Gao, U.~G.~Mei{\ss}ner, Y.~L.~Shen and D.~H.~Li,
Phys. Rev. D \textbf{112}, no.1, 1 (2025)
doi:10.1103/yvjd-2ymn
[arXiv:2412.13084 [hep-ph]].

\bibitem{Huang:2025jsa}
Y.~K.~Huang, D.~H.~Li, C.~D.~L{\"u}, B.~X.~Shi and H.~X.~Yu,
[arXiv:2512.18866 [hep-ph]].

\bibitem{Li:2025mhq}
D.~H.~Li, C.~D.~L{\"u}, U.~G.~Mei{\ss}ner and J.~Gao,
[arXiv:2512.11741 [hep-ph]].

\bibitem{LatticeParton:2022zqc}
J.~Hua \textit{et al.} [Lattice Parton],
Phys. Rev. Lett. \textbf{129}, no.13, 132001 (2022)
doi:10.1103/PhysRevLett.129.132001
[arXiv:2201.09173 [hep-lat]].

\bibitem{BaBar:2009rrj}
B.~Aubert \textit{et al.} [BaBar],
Phys. Rev. D \textbf{80}, 052002 (2009)
doi:10.1103/PhysRevD.80.052002
[arXiv:0905.4778 [hep-ex]].

\bibitem{Belle:2012wwz}
S.~Uehara \textit{et al.} [Belle],
Phys. Rev. D \textbf{86}, 092007 (2012)
doi:10.1103/PhysRevD.86.092007
[arXiv:1205.3249 [hep-ex]].

\bibitem{Ji:2013dva}
X.~Ji,
Phys. Rev. Lett. \textbf{110}, 262002 (2013)
doi:10.1103/PhysRevLett.110.262002
[arXiv:1305.1539 [hep-ph]].

\bibitem{Korchemskaya:1992je}
I.~A.~Korchemskaya and G.~P.~Korchemsky,
Phys. Lett. B \textbf{287}, 169-175 (1992)
doi:10.1016/0370-2693(92)91895-G

\bibitem{Lee:2005gza}
S.~J.~Lee and M.~Neubert,
Phys. Rev. D \textbf{72}, 094028 (2005)
doi:10.1103/PhysRevD.72.094028
[arXiv:hep-ph/0509350 [hep-ph]].

\bibitem{Han:2024fkr}
X.~Y.~Han, J.~Hua, X.~Ji, C.~D.~L{\"u}, W.~Wang, J.~Xu, Q.~A.~Zhang and S.~Zhao,
Phys. Rev. D \textbf{111}, no.11, L111503 (2025)
doi:10.1103/2t8s-w8t6
[arXiv:2403.17492 [hep-ph]].

\bibitem{LatticeParton:2024zko}
X.~Y.~Han \textit{et al.} [Lattice Parton],
Phys. Rev. D \textbf{111}, no.3, 034503 (2025)
doi:10.1103/PhysRevD.111.034503
[arXiv:2410.18654 [hep-lat]].

\bibitem{Ji:2014gla}
X.~Ji,
Sci. China Phys. Mech. Astron. \textbf{57}, 1407-1412 (2014)
doi:10.1007/s11433-014-5492-3
[arXiv:1404.6680 [hep-ph]].

\bibitem{Ji:2020ect}
X.~Ji, Y.~S.~Liu, Y.~Liu, J.~H.~Zhang and Y.~Zhao,
Rev. Mod. Phys. \textbf{93}, no.3, 035005 (2021)
doi:10.1103/RevModPhys.93.035005
[arXiv:2004.03543 [hep-ph]].

\bibitem{Cichy:2018mum}
K.~Cichy and M.~Constantinou,
Adv. High Energy Phys. \textbf{2019}, 3036904 (2019)
doi:10.1155/2019/3036904
[arXiv:1811.07248 [hep-lat]].

\bibitem{Beneke:2023nmj}
M.~Beneke, G.~Finauri, K.~K.~Vos and Y.~Wei,
JHEP \textbf{09}, 066 (2023)
doi:10.1007/JHEP09(2023)066
[arXiv:2305.06401 [hep-ph]].

\bibitem{Liu:2018tox}
Y.~S.~Liu, W.~Wang, J.~Xu, Q.~A.~Zhang, S.~Zhao and Y.~Zhao,
Phys. Rev. D \textbf{99}, no.9, 094036 (2019)
doi:10.1103/PhysRevD.99.094036
[arXiv:1810.10879 [hep-ph]].

\bibitem{Liu:2019urm}
Y.~S.~Liu, W.~Wang, J.~Xu, Q.~A.~Zhang, J.~H.~Zhang, S.~Zhao and Y.~Zhao,
Phys. Rev. D \textbf{100}, no.3, 034006 (2019)
doi:10.1103/PhysRevD.100.034006
[arXiv:1902.00307 [hep-ph]].

\bibitem{Xu:2018mpf}
J.~Xu, Q.~A.~Zhang and S.~Zhao,
Phys. Rev. D \textbf{97}, no.11, 114026 (2018)
doi:10.1103/PhysRevD.97.114026
[arXiv:1804.01042 [hep-ph]].

\bibitem{Ishaq:2019dst}
S.~Ishaq, Y.~Jia, X.~Xiong and D.~S.~Yang,
Phys. Rev. Lett. \textbf{125}, no.13, 132001 (2020)
doi:10.1103/PhysRevLett.125.132001
[arXiv:1905.06930 [hep-ph]].

\bibitem{Zhao:2019elu}
S.~Zhao,
Phys. Rev. D \textbf{101}, no.7, 071503 (2020)
doi:10.1103/PhysRevD.101.071503
[arXiv:1910.03470 [hep-ph]].

\bibitem{Ji:2024oka}
X.~Ji,
Nucl. Phys. B \textbf{1007}, 116670 (2024)
doi:10.1016/j.nuclphysb.2024.116670
[arXiv:2408.03378 [hep-ph]].

\bibitem{Ji:2020brr}
X.~Ji, Y.~Liu, A.~Sch{\"a}fer, W.~Wang, Y.~B.~Yang, J.~H.~Zhang and Y.~Zhao,
Nucl. Phys. B \textbf{964}, 115311 (2021)
doi:10.1016/j.nuclphysb.2021.115311
[arXiv:2008.03886 [hep-ph]].

\bibitem{CLQCD:2023sdb}
Z.~C.~Hu \textit{et al.} [CLQCD],
Phys. Rev. D \textbf{109}, no.5, 054507 (2024)
doi:10.1103/PhysRevD.109.054507
[arXiv:2310.00814 [hep-lat]].

\bibitem{CLQCD:2024yyn}
H.~Y.~Du \textit{et al.} [CLQCD],
Phys. Rev. D \textbf{111}, no.5, 054504 (2025)
doi:10.1103/PhysRevD.111.054504
[arXiv:2408.03548 [hep-lat]].

\bibitem{Tan:2025ofx}
J.~X.~Tan, Z.~C.~Gong, J.~Hua, X.~Ji, X.~Jiang, H.~Liu, A.~Sch{\"a}fer, Y.~Su, H.~Z.~Wang and W.~Wang, \textit{et al.}
Phys. Rev. D \textbf{113}, no.5, 054505 (2026)
doi:10.1103/pry5-7729
[arXiv:2511.22547 [hep-lat]].

\bibitem{Efremov:1979qk}
A.~V.~Efremov and A.~V.~Radyushkin,
Phys. Lett. B \textbf{94}, 245-250 (1980)
doi:10.1016/0370-2693(80)90869-2

\bibitem{Lepage:1980fj}
G.~P.~Lepage and S.~J.~Brodsky,
Phys. Rev. D \textbf{22}, 2157 (1980)
doi:10.1103/PhysRevD.22.2157

\bibitem{Feldmann:2022uok}
T.~Feldmann, P.~L{\"u}ghausen and D.~van Dyk,
JHEP \textbf{10}, 162 (2022)
doi:10.1007/JHEP10(2022)162
[arXiv:2203.15679 [hep-ph]].

\bibitem{Belle:2018jqd}
M.~Gelb \textit{et al.} [Belle],
Phys. Rev. D \textbf{98}, no.11, 112016 (2018)
doi:10.1103/PhysRevD.98.112016
[arXiv:1810.12976 [hep-ex]].

\bibitem{Khodjamirian:2020hob}
A.~Khodjamirian, R.~Mandal and T.~Mannel,
JHEP \textbf{10}, 043 (2020)
doi:10.1007/JHEP10(2020)043
[arXiv:2008.03935 [hep-ph]].

\bibitem{Mandal:2023lhp}
R.~Mandal, S.~Nandi and I.~Ray,
Phys. Lett. B \textbf{848}, 138345 (2024)
doi:10.1016/j.physletb.2023.138345
[arXiv:2308.07033 [hep-ph]].

\bibitem{Han:2024cht}
C.~Han, W.~Wang, J.~L.~Zhang and J.~H.~Zhang,
Phys. Rev. D \textbf{110}, no.9, 094038 (2024)
doi:10.1103/PhysRevD.110.094038
[arXiv:2408.13486 [hep-ph]].

\bibitem{Guo:2025obm}
T.~Guo, C.~Han, W.~Wang and J.~L.~Zhang,
Phys. Rev. D \textbf{112}, no.1, 016013 (2025)
doi:10.1103/c97x-z7j9
[arXiv:2505.08611 [hep-ph]].

\bibitem{Deng:2024dkd}
Z.~F.~Deng, W.~Wang, Y.~B.~Wei and J.~Zeng,
Phys. Rev. D \textbf{110}, no.11, 114006 (2024)
doi:10.1103/PhysRevD.110.114006
[arXiv:2409.00632 [hep-ph]].

\bibitem{Wang:2025uap}
W.~Wang, J.~Xu, Q.~A.~Zhang and S.~Zhao,
Phys. Rev. D \textbf{112}, no.5, 054044 (2025)
doi:10.1103/1547-t91t
[arXiv:2504.18018 [hep-ph]].

\end{thebibliography}
\end{document}